# Integrated machine learning pipeline for aberrant biomarker enrichment (i-mAB): characterizing clusters of differentiation within a compendium of systemic lupus erythematosus patients

Trang T. Le, Ph.D.*[1], Nigel O. Blackwood*[1], Jaclyn N. Taroni, Ph.D.[2], Weixuan Fu, M.S.[1], Matthew K. Breitenstein, Ph.D.[1]

[1]Department of Biostatistics, Epidemiology, and Informatics;
[2]Department of Systems Pharmacology and Translational Therapeutics;
Perelman School of Medicine, University of Pennsylvania, Philadelphia, PA, USA

**Abstract**

Clusters of differentiation (**CD**) are cell surface biomarkers that denote key biological differences between cell types and disease state. CD-targeting therapeutic monoclonal antibodies (**mAB**) afford rich trans-disease repositioning opportunities. Within a compendium of systemic lupus erythematous (**SLE**) patients, we applied the **I**ntegrated **m**achine learning pipeline for **a**berrant **b**iomarker enrichment (**i-mAB**) to profile *de novo* gene expression features affecting CD20, CD22 and CD30 gene aberrance. First, a novel relief-based algorithm identified interdependent features($p$=681) predicting treatment-naïve SLE patients (balanced accuracy=0.822). We then compiled CD-associated expression profiles using regularized logistic regression and pathway enrichment analyses. On an independent general cell line model system data, we replicated associations *(in silico)* of BCL7A($p_{adj}$=1.69e-9) and STRBP($p_{adj}$=4.63e-8) with CD22; NCOA2($p_{adj}$=7.00e-4), ATN1($p_{adj}$=1.71e-2), and HOXC4($p_{adj}$=3.34e-2) with CD30; and PHOSPHO1, a phosphatase linked to bone mineralization, with both CD22($p_{adj}$=4.37e-2) and CD30($p_{adj}$=7.40e-3). Utilizing carefully aggregated secondary data and leveraging *a priori* hypotheses, i-mAB fostered robust biomarker profiling among interdependent biological features.

**Introduction and Background**

***Clusters of differentiation (CD)*** are cell surface biomarkers that denote key biological differences between cell types and disease state. For each of the >400 known CDs [Engel 2015], distinct monoclonal antibodies (**mABs**) enable robust immunophenotyping [Belov 2001, Zucchetto 2011] and serve as scalable biomarkers for translational research [Autenrieth 2015. However, CDs are noticeably modified by upstream, interdependent biological features. Enriching the perspective of CDs holds potential to identify additional novel biomarkers of cell differentiation and activation, and therapeutic repositioning opportunities due to availability of many FDA-approved targeted therapeutic mABs – scalable high-throughput *in silico* approaches are needed to identify interdependent features elucidating the CD landscape.

***B-lymphocytes malignancies and autoimmune disorders.*** B lymphocytes (or B-cells) are white blood cells that are important regulators of the human immune system and function by secreting antibodies, presenting antigen, and secreting cytokines to signal other cells [Pieper 2013]. B-cell dysfunction has wide reaching consequences and can produce a tremendous variety of disease phenotypes, ranging from lymphoma [Ondrejka 2015], autoimmune disorders [Lipsky 2001], and even human immunodeficiency virus pathogenicity [Moir 2008]. This study focuses on the role of B cells in systemic lupus erythematosus (**SLE**), a highly variable, incurable autoimmune disease that can affect any organ system in the human body. SLE is caused by improper B cell behavior, and results in self targeted immune response.

Beyond motivation to elucidate novel CD upstream biology, CD biomarkers hold potential therapeutic repositioning opportunities as many CDs have FDA-approved targeting therapeutic mABs in both B-lymphocyte malignancies [Scott 2012] and autoimmune disorders [Kamal 2014]. Therapeutic mABs can be deployed for antibody-dependent cytotoxicity or as combination therapies enhancing sensitivity to chemotherapy agents [Simpson 2014]. Enriching the perspective of CDs holds potential to elucidate therapeutic repositioning opportunities due to availability of many FDA-approved targeted therapeutic mABs. Scalable high-throughput *in silico* approaches are needed to identify interdependent features elucidating the CD landscape.

Correspondence to: **mkbreit** *at* **pennmedicine** *dot* **upenn** *dot* **edu**

*Machine learning innovations enhance statistical analyses.* We developed the biologically scalable integrated machine learning pipeline for aberrant biomarker enrichment (**i-mAB**) for molecular profiling of the CDs of interest by incorporating multiple recently developed machine learning algorithms. Relief-based algorithms, of which most popular method is ReliefF, are known to effectively capture complex gene-gene interactions that are important for distinguishing classes but often unrecognizable by other algorithms such as Random Forest [Kononenko1997, McKinney 2009]. MultiSURF is an extended version of Relief F that reliably computes significance of features in various data structures including multiple classes with class imbalance [Urbanorwicz 2017]. In updating the feature scores, for a particular observation, while ReliefF considers the same number of nearest neighbors in all classes, MultiSURF automatically computes a neighborhood radius that is flexible throughout the feature space and often contains different number of observations for each class. By adaptively normalizing the weights added to each features based on the proportion of different classes in the neighborhood of each observation, MultiSURF inherently takes into account the class imbalance in the data. We presented a first known application of the novel Relief-based algorithm MultiSURF to real-world biomedical data to identify the most predictive gene expression features in classifying patients and quantify their predictive power with the automated machine learning system Tree-based Pipeline Optimization Tool (TPOT) [Olson 2016]. Using genetic programming, TPOT optimizes a series of feature preprocessing techniques and machine learning models and searches for the best prediction pipeline of different machine learning operators with tuned hyperparameters. We used TPOT to obtain the optimal framework for the training data with the objective of maximizing the cross-validated balanced accuracy and reported the out-of-sample balanced accuracy for classifying the patient groups in the testing data.

*Study motivation.* The goal of the current study was to utilize i-mAB to enrich the perspective of CDs with interdependent gene expression features and identify novel upstream transcriptomic biomarkers that characterize aberrance of CD20, CD22, and CD30 expression. In particular, our study sought to enrich the perspectives of CD20 (*MS4A1* - Membrane-spanning 4-domains subfamily A member 1), CD22 (*SIGLEC2* - Sialic acid-binding Ig-like lectin 2), and CD30 (*TNFRSF8* - Tumor necrosis factor receptor superfamily member 8), due to characteristic overexpression in both B-lymphocyte hematologic malignancies and autoimmune disorders. We exclusively focused within gene expression characterization, for the purpose of evaluating aberrant CD expression. By incorporating clearly defined hypotheses with machine learning applications robust to multi-collinearity, we aim to enrich our perspective of CD biology and potentially leading trans-disease therapeutic repositioning opportunities.

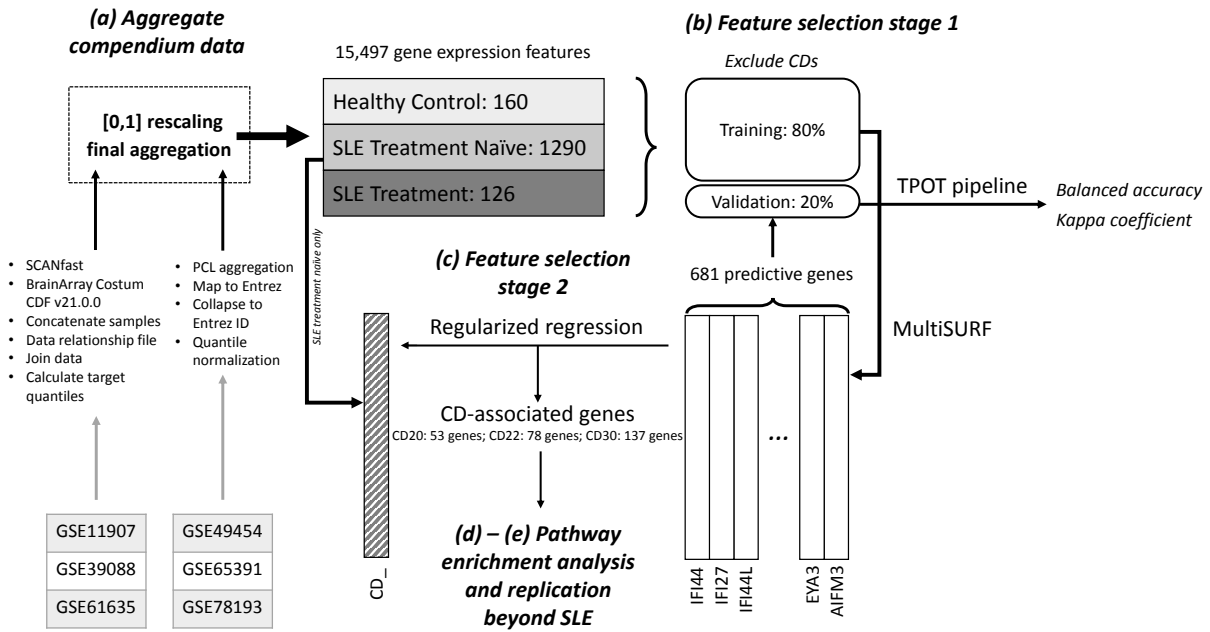

**Figure 1. Integrated machine learning pipeline for aberrant biomarker enrichment (i-mAB) study overview.** (a) Compendium data assembly: Preprocess and aggregation of gene expression data from six different studies resulting in a compendium of 160 healthy samples, 1290 SLE samples treatment naïve, and 126 SLE samples with treatment. (b) Features selection stage 1: identifying predictive genes in classifying patients with SLE treatment naïve

using MultiSURF. (c) Feature selection stage 2: detecting genes associated with aberrant level of CDs using regularized logistic regression. (d)-(e) Pathway enrichment analysis and replication beyond SLE cohort.

**Methods**

*Study Overview:* In this study, we aimed to detect and characterize CDs-related genes among the most predictive genes in classifying the samples into three categories: healthy, SLE treatment naïve, and SLE treatment. For editorial clarity, the CD nomenclature [Engel 2015] will be deployed to reference gene expression of CD20, CD22, and CD30 within methods and results section will reference gene expression, as opposed to the more ubiquitous deployment of CD nomenclature in diagnostic proteomics. After gathering the most important features (genes) for the classifier, we used regularized logistic regression to robustly identify genes whose expression is either convergent or divergent in contribution to the effect on the aberrant expression of the CDs of interest within the SLE treatment naïve group. We also performed a pathway enrichment analysis of these genes to gain further insights into their biological and functional characteristics. An overview of compendium assembly and i-mAB pipeline are shown in **Figure 1**; i-mAB packages are provided on the Breitenstein Lab GitHub page: https://breitensteinlab.github.io/i-mAB/

*(a) SLE Compendium assembly.* A compendium of health controls, treatment naïve SLE patients, and SLE patients exposed to various treatments was assembled using data from Gene Expression Omnibus [Barrett 2012] – representing our *'SLE Compendium'*. (**Note**: a subset of patients within the 'treatment naïve' group received maintenance immunosuppressive therapy, but were not exposed experimental treatments). This compendium encompassed human-derived gene expression measures from 6 original studies, including: GSE11907 [Chaussabel 2008], GSE39088 [Lauwerys 2013], GSE72747 [Ducreux 2016], GSE49454 [Chiche 2014], GSE61635 [Carpintero 2015], GSE65391 [Banchereau 2016], GSE78193 [Welcher 2015]. Our study exclusively utilized existing, de-identified data from human subjects and did not require local Institutional Review Board review. Affymetrix data was processed with Single Channel Array Normalization (SCAN) [Piccolo 2012]. Other platform data (*e.g.*, Agilent, Illumina) were quantile normalized using the Affymetrix data as a reference. Each individual dataset was scaled from 0 to 1 on a per-gene basis before concatenating the data sets. Detailed data preprocessing and aggregation steps (including source code) are available at https://github.com/greenelab/rheum-plier-data. Detailed sample characteristics of the SLE Compendium can be found as supplementary publication [Le 2018].

*(b) Feature selection stage 1: identifying predictive genes in classifying patients with SLE treatment naive.*
*MultiSURF-guided feature inclusion.* The dataset was randomly split into 80% for training and 20% for validation. On the training samples, we applied MultiSURF to obtain feature importance scores and extracted the $p$ most predictive features (genes) that were input of the second stage of the analysis. We remarked that the rescaling of the importance scores to range from -1 to 1 does not affect the relative importance among features. To prevent overfitting, we excluded all known CDs from this first analysis step.

*Predictive power estimation with TPOT.* In order to quantify the classification accuracy provided by the MultiSURF features, we applied TPOT on training samples to get the optimized prediction pipeline and then implemented the pipeline on the training set with iterative inclusion of the features with highest MultiSURF importance scores and reported the pipeline's performance on the testing set. In other words, we assessed the predictive power of the $p$ features by applying the recommended set of operators with increasing number of features to obtain predictions of patient types in the validation set. Considering the class imbalance in our compendium data, in order to properly evaluate the performance of each model, we calculated the values of standard accuracy (proportion of correct predictions), balanced accuracy (mean of sensitivity and specificity), and Kappa coefficient which is an accuracy measure that is scaled to expected accuracy.

When performing feature selection among genes sampled at multiple time points, there is the potential for autocorrelation of the expression of genes over time. However, our goal was to find a comprehensive list of gene expression features that might affect CD expression levels regardless of the sampled time point. Thus, we treated a given transcript's expression at each time point as independent of other time points. This increases the power to detect multiple time-dependent responses. For example, a certain gene might be important because of its role in early response to treatment whereas another gene might be activated in a later secondary response. Choosing only one time-point or averaging over time would decrease the power detect a variety of such gene expression signals. This is also the reason we did not implement popular techniques to treat data's imbalance such as resampling or generating artificial samples, which do not reduce the bias toward the majority class in high dimensional data [Lusa 2013]. We remark that, despite having multiple time samples, some effects still might be difficult to detect because some individual's genes might peak at a different characteristic time point for a given cellular response.

*(c) Feature selection stage 2: detecting CD-related genes with regularized logistic regression.* Within our SLE Compendium, gene expression of all known CDs, including CD20, CD22 and CD30, were categorized as 'aberrant' or 'non-aberrant' based on the following criteria: *i) two-tailed normalization* at 20th and 80th percentile of relative gene expression. The two tails encompassed 'aberrant' CD expression, whereas the middle distribution served as 'non-aberrant'. Based on histogram distributions, threshold adjustment were necessary for a subset of CDs: *ii) Adjusted two-tail normalization* (n=4) was applied when CD expression followed an apparent normal distribution, but the default thresholds did not satisfactorily characterize feature variation, or *iii) binarization of non-normal distributions* (n=86) separated the expression values into low/high groups (instead of non-aberrant/aberrant) to characterize apparent patterns of CD expression distributions. (**Note**: two-tailed normalization without adjustment was applied for CD20, CD22, and CD30). Detailed labelling of all CD features (histograms of gene expression distribution, descriptive statistics of overall expression and variation within the SLE Compendium) are provided within a supplementary manuscript [Le 2018].

Elastic net was chosen for regularization of gene expression features in our second feature selection stage due to known robustness within bioinformatics applications using high-dimensional data with highly correlated biological features [Zou 2005]. We performed elastic net regularized logistic regressions on each of the categorized CD variable and identified a set of $k$ features ($k < p$) that are associated with each CD expression among the $p$ previously selected features predicting SLE treatment naïve patients (*feature selection stage 1)*. Incorporating Lasso (L1) and ridge regression (L2) penalties, the elastic net simultaneously selects variables and shrinks the coefficients of correlated predictors. We set the hyper-parameter α=0.5 to balance the proportion of L1 and L2 penalty and tuned the regularization parameter λ with cross validation to obtain the best model containing the genes that are associated with the expression level of the CD of interest. Notably, elastic net will tend to give strongly correlated genes similar regression coefficients. These genes were then ranked based on the adjusted *p*-value resulting from their independent logistic regression of the CD Aberrant/Non-aberrant expression groups. Independent odds ratio for each association was also reported. Further, because data on gender and age are not available for two of the six studies, we did not correct for these covariates to preserve the power of the analysis.

*(d) Feature annotation: pathway enrichment analysis of CD-specific gene expression profiles.* Gene Set Enrichment Analysis (GSEA) is an open-access software that computes the degree of overlap between a predefined gene set and collection of annotated gene sets in the Molecular Signatures Databases (**MSigDB**) [Subramanian 2005]. We use this tool to search for enriched Reactome and molecular function pathways among the CD-associated genes.

*(e) Independent in silico replication in general cell line model systems.* A panel 64 human-derived general cell line models, measuring 12,073 gene expression features, from the Human Cell Atlas [Thul 2017] served as independent *in silico* replication. (https://www.proteinatlas.org/download/rna_celline.tsv.zip) We performed a correlation test of the counts in each specific cell types sample between that gene and its corresponding CD expression on features identified during *feature selection stage 2*.

**Results**

*(a) Assembly of SLE Compendium.* Our compendium of human SLE patients contained 1,576 observations, with multiple measures per patient, aggregated from original studies [Chaussabel 2008, Lauwerys 2013, Ducreux 2016, Chiche 2014, Carpintero 2015, Banchereau 2016, Welcher 2015]. The SLE compendium contained 15,497 gene expression measurements with observations from healthy control (n=160) samples, treatment-naïve SLE (n=1,290) samples, and SLE samples exposed to various treatments (n=126) (**Table 1**).

*(b) Feature selection stage 1 – Gene expression profile of treatment naïve SLE patients.* In our study, maximizing prediction balanced model accuracy was only a minor component of our gene expression profiling, with maximizing opportunity for biologically rich and inferential signals being of most importance. Further complicating gene expression profiling endeavors was the known issue possibility of multicollinearity, where many biologically important signals are correlated with both other explanatory features and the study outcomes. Therefore, selection of the mathematically robust model MultiSURF with an inclusive, albeit replicable, feature inclusion threshold of 0.177 (1500/max(raw feature score)) was chosen based on the distribution of the importance scores [**Supplement 1. Figure S1**]. This heuristic threshold yields a reasonable number of genes for the next step of the analysis. Applying this threshold, we collected *p*=681 gene expression features that have significantly high total importance score compared to the remaining genes. We reiterate that rescaling the MultiSURF importance scores to range from -1 to 1 does not affect the relative importance among features.

**Table 1. SLE Compendium characteristics as ascertained from study of origin**

|  | Cohort 1 | Cohort 2 | Cohort 3 | Cohort 4 | Cohort 5 | Cohort 6 | Overall |
|---|---|---|---|---|---|---|---|
| **Study PMID** | 18631455 | 23203821 | 24644022 | 25736140 | 27040498 | 26138472 | --- |
| **Study GEO identifier** | GSE11907 | GSE39088 | GSE49454 | GSE61635 | GSE65391 | GSE78193 | --- |
| **Healthy control*** | 0 | 46 | 0 | 30 | 72 | 12 | 160 |
| median age (range) | --- | 34.5 (19-50) | --- | --- | 12 (6-21) | --- | 16 (6-50) |
| gender - female/male | --- | 34 | --- | --- | 57 | --- | 91 |
| **SLE-treatment naïve*** | 37 | 21 | 177 | 99 | 924 | 32 | 1290 |
| median age (range) | 14 (8-17) | 43 (20-50) | 40 (18-71) | --- | 15 (6-19) | --- | 16 (6-71) |
| gender - female | 35 | 21 | 148 | --- | 817 | --- | 1021 |
| **SLE-various treatments*** | 0 | 57 | 0 | 0 | 0 | 69 | 126 |
| median age (range) | --- | 36 (19-50) | --- | --- | --- | --- | 36 (19-50) |
| gender - female/male | --- | 57 | --- | --- | --- | --- | 57 |

*observation characteristics represent multiple observations per patient*

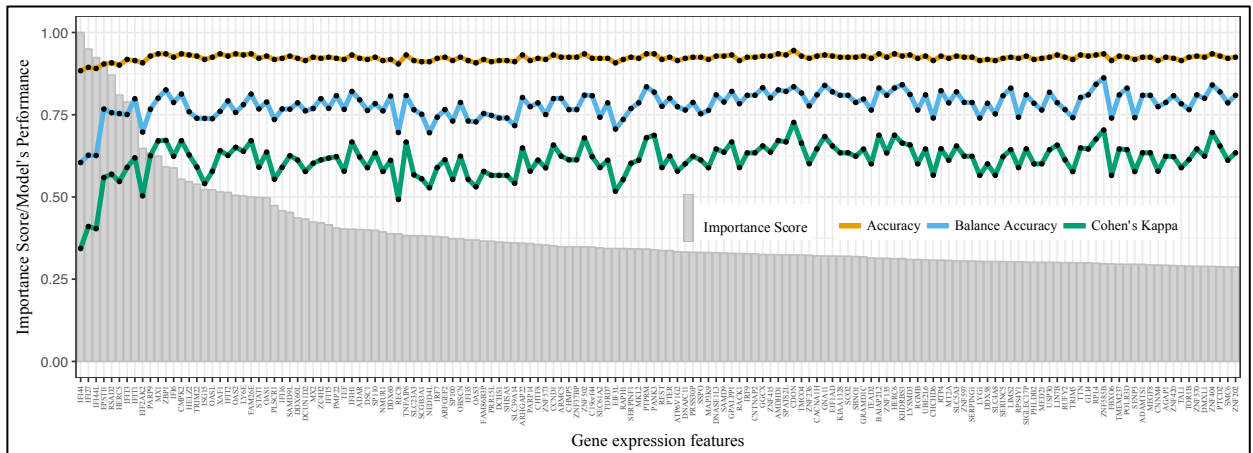

**Figure 2**. Gene expression feature importance profile of treatment naïve systemic lupus erythematosus patients *(top 100 features listed)*. Results from *step b*, containing 681 gene expression features that differentiate treatment naïve SLE patients from healthy controls and SLE patients exposed to various treatment. Feature importance profile includes: i) scaled importance score (grey bars) and ii) corresponding out-of-sample classification accuracy of sample type (HC/SLE treatment naïve/SLE treatment) from adjusted TPOT-recommended pipeline with iterative inclusion of features from left to right (orange, blue or green lines). The y-axis represents both the MultiSURF scaled importance score and TPOT pipeline accuracy.

We noted that our focus at the first stage of the analysis is feature inclusion; therefore, we only reported the performance from the optimized pipeline as an estimation of the model's predictive power. TPOT suggested a complex pipeline that stacks the gradient boosting, decision tree and Random Forest algorithm with an intermediate step of selecting the top 20-percentile features based on their ANOVA F-values between with the class. We adjusted the TPOT-recommended pipeline slightly by removing one step of feature selection in order to obtain the corresponding out of sample accuracy as we include more features in the stacked model (**Figure 2**). Initially, as more predictors were included in the model, the out of sample accuracy increased, demonstrating that the added features are meaningful. However, we note that after the inclusion of approximately ten most predictive features in the model, the increase in balanced accuracy (yellow) and Kappa coefficient (blue) slowed down. Nevertheless, there was an overall upward trend in these performance metric values as more features were added to the model. We also noted that the flow of balanced accuracy and Kappa coefficient are not smooth, which was likely due to the built-in stochasticity of the model. To prevent any biases in the feature scoring metric of the algorithm, we consider all 681 genes for the second stage of finding association with CD expression. We note that the 681-predictor model attains a relatively high out-of-sample prediction accuracy of approximately 0.935, balanced accuracy of 0.822, and Kappa coefficient of 0.680. We recalled the Kappa coefficient is an accuracy measure that is scaled to expected accuracy which is the random chance of making a correct prediction by a null model. In particular, a Kappa coefficient of 0.680 means that the model achieved a rate of classification 68% of the way between a null model and perfect classification.

*(c) Feature selection stage 2 – CD-specific gene expression profiles.* Within the 681 features selected during *(b) feature selection stage 1*, elastic net regularized binomial logistic regression model to select features that are statistically associated with select CDs and calculated the binomial deviance *D*, the conventional measure of the lack of fit to the data in a logistic regression model. Fixing the hyper-parameter α=0.5, we used cross-validation to tune the regularization parameter λ. Overall, the regularized logistic regression achieved high correlation with CD20 (*D*=0.0698, λ=0.0335, *k*=53 features), CD22 (*D*=0.1351, λ=0.0310, *k*=78 features), and CD30 (*D*=0.1908, λ=0.0221, *k*=137 features). Selected gene expression features were then characterized as univariate associations (*i.e.* independent effects) on the same CD endpoints. Gene expression features, in addition to corresponding effect size and significance, varied widely between CD20, CD22, and CD30 (**Figure 3**). Odds ratios and 95% confidence intervals characterize increasing explanatory gene expression unless designated otherwise (*=decreasing explanatory gene expression). Even though independent analyses do not reveal the significance of several genes (95% CI of odds ratio well contains the null value of 1, such as MARCKSL), they are kept in the elastic net algorithm due to their contribution to the amount of variance explained in the regression model. We note that the odds ratios and their 95% CI are shown without including study origin as a covariate due to relative consist distribution of CD expression across the compendium studies [**Supplement 2. Figure S2**]. However, we performed additional regressions to explore a potential study of origin effect, and only the regression of CD30 aberrant expression level suggested a potential difference between GSE49454/GSE61635 and the remaining studies [**Supplement 3. Figure S3**]. For consistency, we showed the results from simple regressions with only one explanatory variable (gene expression).

*(d) Feature annotation: pathway enrichment analysis of CD-associated genes.* We performed GSEA of molecular function and biological processes amongst features recommended by *feature selection stage 2* to enhance our *de novo* profile with existing knowledge bases. Our motivation was two-fold: *i)* We recognized that by study design, *feature selection stage 1 – gene expression profile of treatment naïve SLE patients* had potential to introduce false negatives into CD-specific profiles due to reduction of the feature space, and *ii)* while biological activity typically consists of tightly-connected reactions and interactions, statistical signals might be too disparate to clearly resonate within existing biological knowledge.

Our GSEA identified several noteworthy findings within biological processes: **Phosphate-containing compound metabolic process** was identified for CD22 (k/K=0.0076, $1.94e^{-3}$, encompassing: DLG1, DUSP15, EPHB4, IKBKAP, MAP2K6, MSH2, NDUFB1, NUDT5, PDE8A, PDGFB, PHOSPHO1, RFK, TNK2, TTN, and TYMP) and CD30 (k/K=0.0086, $1.53e^{-2}$, encompassing: ABHD14B, IRS1, ISYNA1, MAP2K6, ME1, NDUFB1, OBSCN, PANK3, PDE8A, PDGFB, PHOSPHO1, PI4K2A, PRKD3, PSMB4, RIPK3, SMPD3, and TNK2). The closely-related **organophosphate metabolic process** was also identified for CD22 (k/K=0.0076, $1.94e^{-3}$, encompassing: DLG1, MSH2, NDUFB1, NUDT5, PDE8A, PDGFB, PHOSPHO1, RFK, TYMP) and CD30 (k/K=0.0076, $1.94e^{-3}$, encompassing: ABHD14B, IRS1, ISYNA1, ME1, NDUFB1, PANK3, PDE8A, PDGFB, PHOSPHO1, PI4K2A, and SMPD3). **Kinase activity**, catalysis a phosphate group to a substrate molecule, for CD22 (k/K=0.0083, q=$3.64e^{-2}$, encompassing: ACSL6, ABCB4, ATN1, BAG2, CEP68, EIF3L, IKBKAP).

For CD20, several signals broadly encompassing tissue development and function were identified, including: **muscle contraction** (k/K=0.0172, q=$4.79e^{-2}$), **muscle organ development** (k/K=0.0181, q=$2.57e^{-2}$), **muscle structure development** (k/K=0.0116, q=$4.79e^{-2}$), **muscle system process** (k/K=0.0177, q=$2.57e^{-2}$), **organ morphogenesis** (k/K=0.0083, q=$3.80e^{-2}$). **Calmodulin binding**, implicating intracellular calcium receptor regulation, was also linked to CD20 (k/K=0.0279, q=$1.14e^{-3}$, encompassing: SCN5A, USP6, TTN, MYH3, MARCKSL1). Calmodulin affects a wide range of physiological processes, including cell proliferation, apoptosis, autophagy, and cancer cell differentiation [Berchtold 2014]. Detailed associations are provided for the **cellular function [Supplement 4. Table S1.]** and **molecular pathways [Supplement 5. Table S2.]** gene enrichment analyses.

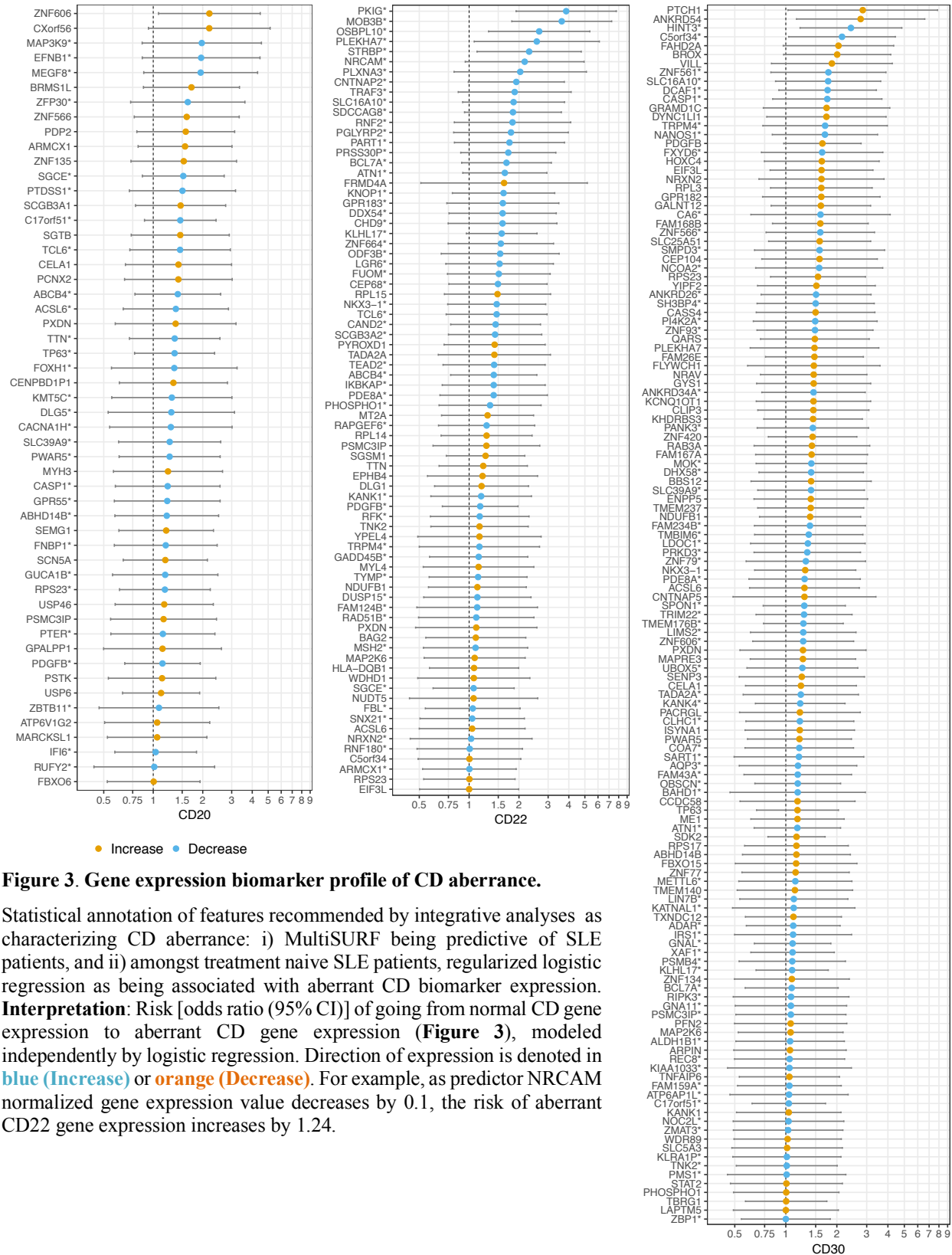

**Figure 3**. Gene expression biomarker profile of CD aberrance.

Statistical annotation of features recommended by integrative analyses as characterizing CD aberrance: i) MultiSURF being predictive of SLE patients, and ii) amongst treatment naive SLE patients, regularized logistic regression as being associated with aberrant CD biomarker expression. **Interpretation**: Risk [odds ratio (95% CI)] of going from normal CD gene expression to aberrant CD gene expression (**Figure 3**), modeled independently by logistic regression. Direction of expression is denoted in blue (Increase) or orange (Decrease). For example, as predictor NRCAM normalized gene expression value decreases by 0.1, the risk of aberrant CD22 gene expression increases by 1.24.

*(e) Independent in silico replication of CD-specific gene expression profile in general cell model systems.* We performed a correlation test between the transcript counts of the detected CD-associated genes with the corresponding CD in 64 human-derived cell lines. Among 78 gene expression features that were previously selected by elastic net to be associated with aberrant level of CD22, we found three genes, BCL7A, STRBP and PHOSPHO1, that have statistically significant correlation with the CD22 expression level in the Human Protein Atlas cell line database, after adjusting the p-values with the Benjamini-Hochberg's procedure [Benjamini 2001]. For CD30, among 137 gene expression features that were identified by elastic net, four genes were shown to significantly correlate with the CD expression level: NCOA2, PHOSPHO1, ATN1, and HOXC4. None of the 53 CD20-related genes showed significant correlation in this cell line database after p-value correction.

**Table 2. *In silico* replication of features affecting CD aberrance in human-derived cell line models**

|  | Gene | Correlation | (95% CI) | t-statistic | Unadjusted p-value | Adjusted p-value |
|---|---|---|---|---|---|---|
| CD22 | *BCL7A* | 0.719 | (0.575, 0.820) | 8.14 | $2.25e^{-11}$ | $1.69e^{-9}$ |
|  | *STRBP* | 0.672 | (0.510, 0.787) | 7.14 | $1.23e^{-9}$ | $4.63e^{-8}$ |
|  | *PHOSPHO1* |  | (0.152, 0.575 | 3.27 | $1.75e^{-3}$ | $4.37e^{-2}$ |
| CD30 | *NCOA2* | 0.534 | (0.331, 0.689) | 4.97 | $5.58e^{-6}$ | $7.00e^{-4}$ |
|  | *PHOSPHO1* | 0.464 | (0.246, 0.637) | 4.12 | $1.13e^{-4}$ | $7.40e^{-3}$ |
|  | *ATN1* | 0.430 | (0.206, 0.611) | 3.75 | $3.88e^{-4}$ | $1.71e^{-2}$ |
|  | *HOXC4* | 0.401 | (0.173, 0.589) | 3.45 | $1.01e^{-3}$ | $3.34e^{-2}$ |

**Discussion**

Using the i-mAB pipeline we identified several noteworthy findings that enrich our understanding of CD biology: For CD22 and CD30 we found phosphate-containing compound metabolic process, organophosphate metabolic process, kinase activity (phosphate catalysis). In independent *in silico* replication, *PHOSPHO1*, a phosphatase linked to bone mineralization, was associated with both CD22 and CD30 gene expression. Future *in vitro* study is warranted to elucidate the potential implication of phosphates on aberrant CD22 and CD30 expression. For CD20, several signals broadly encompassing tissue development and function were identified, including: muscle contraction muscle organ development, muscle structure development, muscle system process, organ morphogenesis, and most interestingly, calmodulin binding (intracellular calcium receptor regulation).

*Strength and limitation.* Aggregation of public datasets may provide a number of advantages but also disadvantages including potential bias due to study origin. However, we closely followed the guideline suggested by Smith et al. in analyzing secondary datasets to produce meaningful results [Smith 2011] and paid close attention to the aggregation procedure to minimize potential bias across studies. We utilized robust methodologies, careful selected models, and sound analytical comparisons to identify gene expression signals with potential for biological relevance. In our study, two key methodological considerations existed, encompassing three parts: i) initial gene expression feature generation (*i.e.* feature selection stage 1) representative of treatment naive SLE patients and ii) attribution of features (*i.e.* feature selection stage 2) to CDs as gene expression profiles, and iii) feature annotation. i) In pursuit of developing an agnostic gene expression profile of treatment naive SLE patients, we were required to make imbalanced comparisons between healthy normal controls (*n*=160) and treated SLE patients (*n*=126) to treatment naive SLE patients (*n*=1,290). However, we took a series of steps to overcome potential limitations due to imbalanced comparisons. First, MultiSURF, a feature selection method known to be robust to imbalanced data, served as our agnostic feature generator and identified gene expression features of potential relevance. Second, an automated machine learning system recommended an optimized pipeline with multiple complex algorithms that would not have been implemented manually without automated machine learning. We focused on the completeness of measures of the model's performance and reported the Cohen's Kappa coefficient as well as balanced accuracy while considering the multi-class and imbalanced-class problems. ii) From a statistical perspective, gene expression can be highly correlated potentially because the perspective of the aggregate transcriptome might be indiscriminate to complex biological synthesis and regulatory pathways influencing gene expression. We applied a regularized multivariate logistic regression to identify the predictive features that are statistically significantly associated with the aberrant level of CDs expression while taking into account the data's multicollinearity. iii) Univariate associations between individual gene expression features and select CDs were independently tested. However, considering the interaction among the genes, we focused on the pathway enrichment analysis of CD-associated genes. We highlighted certain single gene expression features due to transparent biological relevance and confidence in strength of signal. Further, the independent in silico replication of the several CD-specific gene expression profiles in general cell model systems also ascertained the association between these genes and the CD expression. Incorporating *a priori* hypotheses in newly developed, data-driven machine learning methods, i-mAB provides a biologically scalable pipeline for profiling CDs and potentially other interdependent biomarkers such as cytokines.

*Consideration for generalizability*. Some protein products degrade rapidly while others are persistent for a long time. Similarly, transcripts are known to have dramatic variation in persistence - elevated transcript levels might be necessary to produce comparable levels of bioavailable protein products in comparison to more stable proteins (*e.g.* proteins within the same pathways). Aberrant biology occurring within the SLE disease state has potential to indeterminately modify transcript synthesis or protein bioavailability (*i.e.* stability, degradation) and maintenance of physiological homeostasis. A single statistical or machine learning approach often provides a wide-angle view of the biological picture of disease. Careful iterative analysis with multiple approaches may provide a higher resolution picture of complex mechanism and signals of disease. While the signals replicated within general cell model systems potentially suggest broader biological implications, these biomarkers might have limited application to whole blood and B-lymphocytes. As previously noted, a subset of patients labeled as treatment naïve likely received maintenance immunosuppressive therapy. Although the therapy was not an active treatment for the disease, it may have slightly attenuated the overall transcriptome expression. Replication of the pathway findings and independent replication signals are warranted for different diseases and tissue-specific environments

*Potential biomarker applications of upstream biological features as potential biomarkers.* CDs may represent biological relevant markers of disease. Due to orphan drug policy [Braun 2010], enriched CD perspectives might stimulate opportunities for therapeutic repositioning across disease with similar biomarker expression. Particularly for treatment of rare diseases [Seoane 2008] therapeutic mABs have been previously demonstrated to be safe in humans.

**Conclusion**

The i-mAB pipeline identified novel (adjusted independent) associations of potential relevance to CD biology: BCL7A (p=1.69e-9) and STRBP (p=4.63e-8) with CD22; NCOA2(p=7.00e-4), ATN1(p=1.71e-2), and HOXC4 (p=3.34e-2) with CD30. PHOSPHO1, a phosphatase linked to bone mineralization, was associated with both CD22 (p=4.37e-2) and CD30 (p=7.40e-3) expression. Simultaneously leveraging *a priori* hypotheses, performing secondary data analysis, and integrating appropriate machine learning approaches, i-mAB provides opportunity to detect *de novo* gene expression features that replicate in independent disease-agnostic model systems and enrich our understanding of the molecular characteristics of SLE and select CDs.

**Supplement 1. Figure S1. Sorted MultiSURF feature importance scores of all gene expression features.**

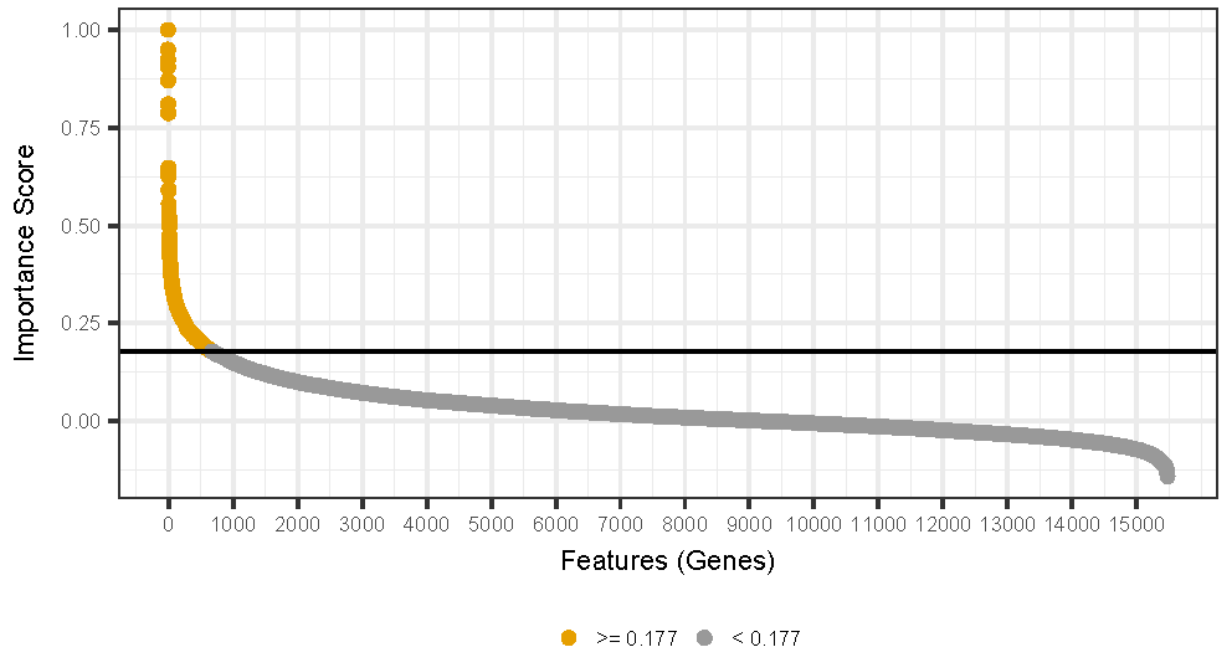

**Figure S1**. Feature importance threshold of 0.177 was selected heuristically based on the distribution of scores. The 681 gene expression features (**yellow**) above this threshold were selected for inclusion within the gene expression profile of SLE, treatment naïve patients.

**Supplement 2. Figure S2. Distribution of CD20, CD22, & CD30 gene expression by original study cohort**

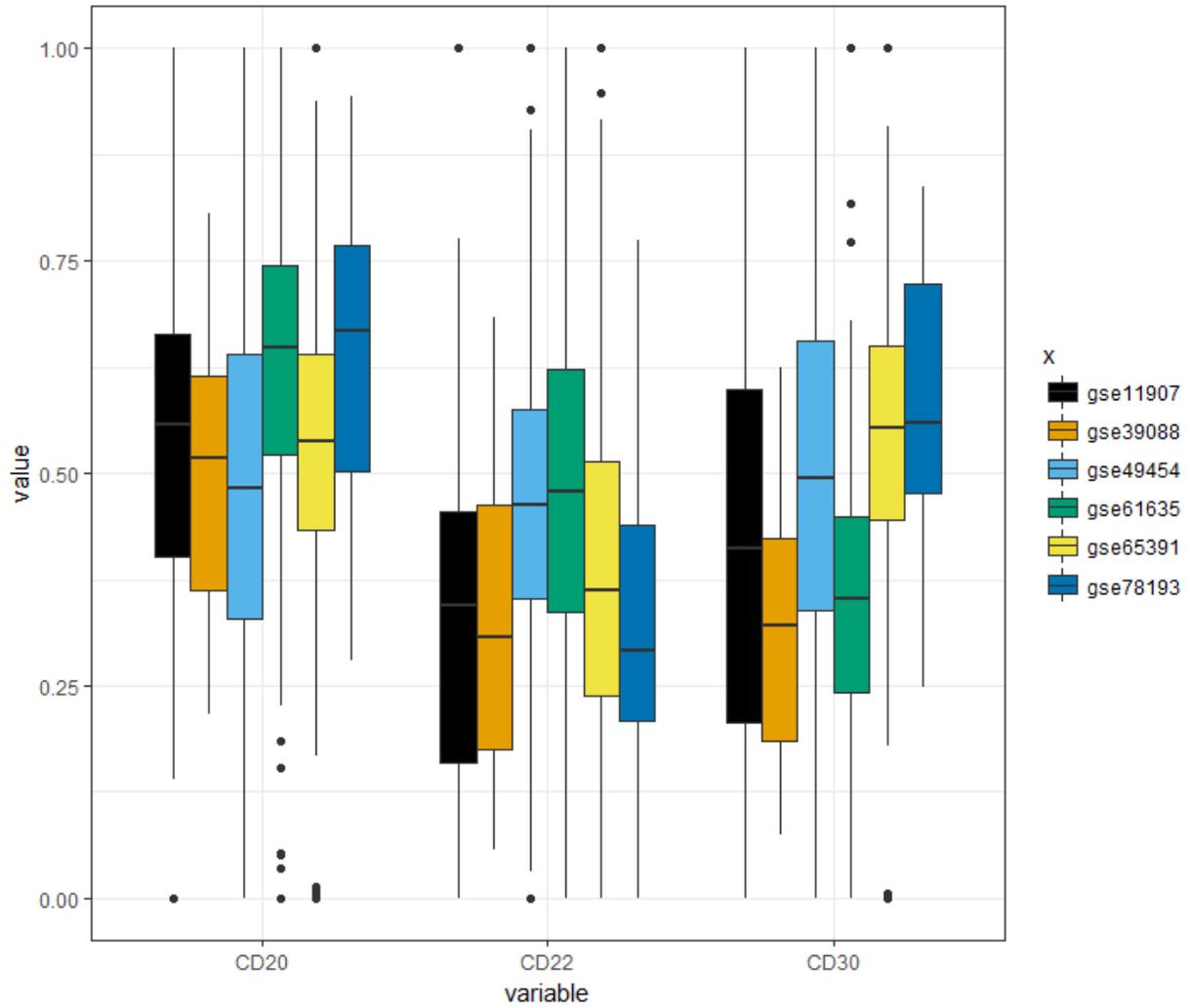

**Figure S2.** Boxplots and 95% confidence intervals of normalized gene expression for CD20, CD22, and CD30 on a scale of 0 – 1. Only minor variation was observed between original study cohorts.

**Supplement 3. Figure S3. Batch effect characterization for CD20, CD22, & CD30**

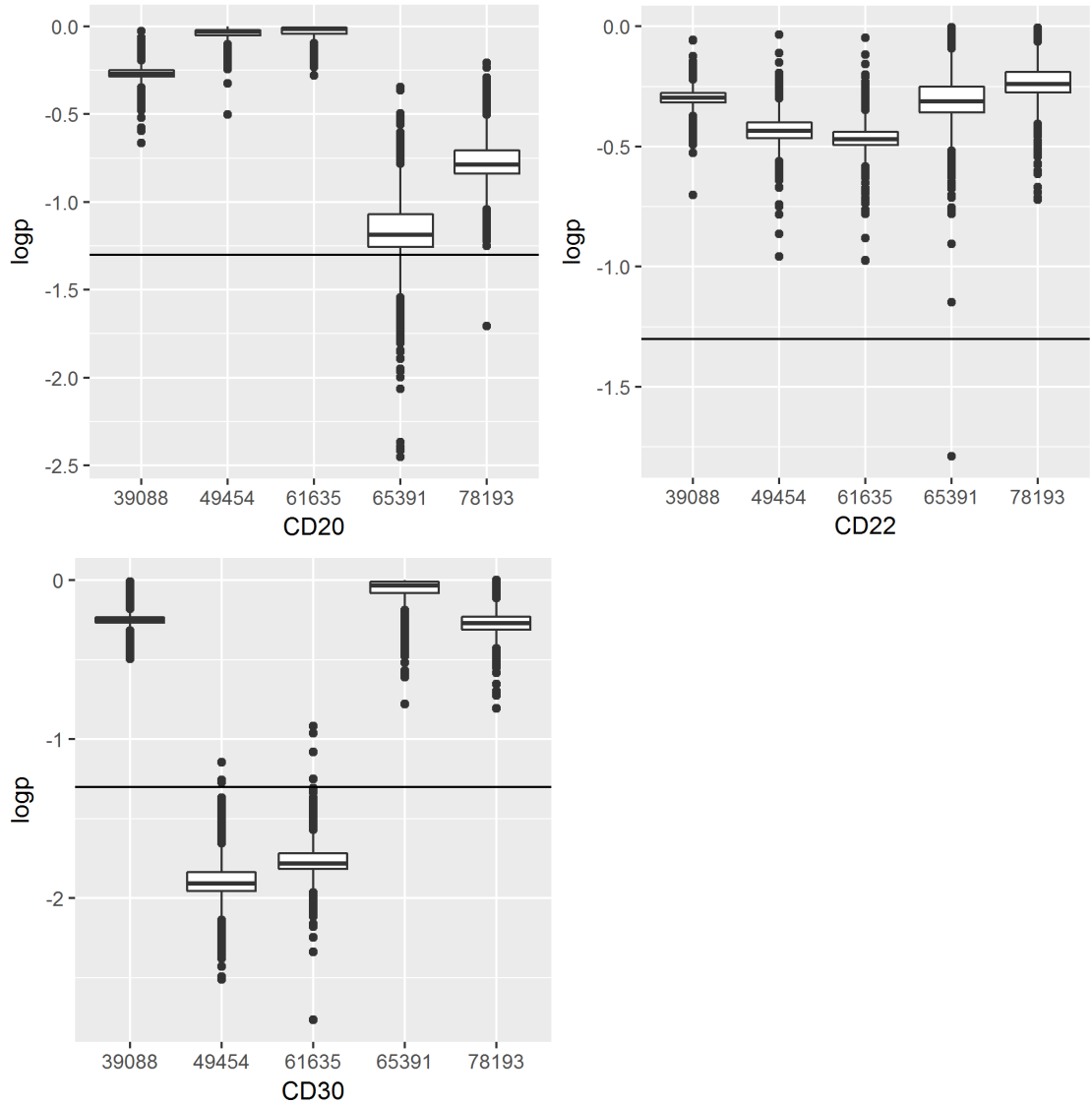

**Figure S3.** Boxplots and 95% confidence intervals of log10 p-values comparing original study cohort (GSE ID provided) to GSE11907. A potential batch (cohort) effect was denoted for GSE49454 and GSE61635 for CD30.

**Supplement 4. Table S1. Cellular function gene sets.**

| CD Predictor | Gene Set Name [1] | Gene Set Description | k / K [2] | p-value [3] | FDR [4] q-value | Genes contained within set overlap |
|---|---|---|---|---|---|---|
| CD20 | Actin-mediated cell contraction | The actin filament-based process in which cytoplasmic actin filaments slide past one another resulting in contraction of all or part of the cell body. | 3 / 74 | 6.58E-05 | 4.79E-02 | SCN5A, TTN, MYH3 |
| CD20 | Enzyme linked receptor protein signaling pathway | Any series of molecular signals initiated by the binding of an extracellular ligand to a receptor on the surface of the target cell, where the receptor possesses catalytic activity or is closely associated with an enzyme such as a protein kinase, and ending with regulation of a downstream cellular process, e.g. transcription. | 6 / 689 | 7.98E-05 | 4.79E-02 | FOXH1, MEGF8, PDGFB, EFNB1, ATP6V1G2, GUCA1B |
| CD20 | Glycoprotein catabolic process | The chemical reactions and pathways resulting in the breakdown of glycoproteins, any protein that contains covalently bound glycose (i.e. monosaccharide) residues; the glycose occurs most commonly as oligosaccharide or fairly small polysaccharide but occasionally as monosaccharide. | 2 / 15 | 1.11E-04 | 4.79E-02 | CELA1, FBXO6 |
| CD22 | Immune response | Any immune system process that functions in the calibrated response of an organism to a potential internal or invasive threat. | 11 / 1100 | 1.48E-06 | 2.00E-03 | DLG1, GPR183, HLA-DQB1, IKBKAP, MSH2, MT2A, PGLYRP2, PXDN, TNK2, TRAF3, TRPM4 |
| CD22 | Immune system process | Any process involved in the development or functioning of the immune system, an organismal system for calibrated responses to potential internal or invasive threats. | 14 / 1984 | 2.98E-06 | 2.00E-03 | ABCB4, DLG1, GPR183, HLA-DQB1, IKBKAP, MAP2K6, MSH2, MT2A, PDGFB, PGLYRP2, , PXDN, TNK2, TRAF3, TRPM4 |
| CD30 | Intracellular signal transduction | The process in which a signal is passed on to downstream components within the cell, which become activated themselves to further propagate the signal and finally trigger a change in the function or state of the cell. | 15 / 1572 | 2.82E-05 | 1.53E-02 | GPR182, IRS1, MAP2K6, NKX3-1, PDGFB, PRKD3, PSMB4, RAB3A, RIPK3, SART1, STAT2, TMBIM6, TNK2, TP63, ZMAT3 |
| CD22 | Mechanosensory behavior | Behavior that is dependent upon the sensation of a mechanical stimulus. | 3 / 12 | 8.73E-07 | 1.94E-03 | CNTNAP2, NRXN2, STRBP |
| CD20 | Membrane depolarization | The process in which membrane potential decreases with respect to its steady-state potential, usually from negative potential to a more positive potential. For example, the initial depolarization during the rising phase of an action potential is in the direction from the negative steady-state resting potential towards the positive membrane potential that will be the peak of the action potential. | 3 / 61 | 3.69E-05 | 4.09E-02 | SCN5A, CACNA1H, CASP1 |
| CD22 | Membrane organization | A process which results in the assembly, arrangement of constituent parts, or disassembly of a membrane. A membrane is a double layer of lipid molecules that encloses all cells, and, in eukaryotes, many organelles; may be a single or double lipid bilayer; also includes associated proteins. | 10 / 899 | 1.81E-06 | 2.00E-03 | ABCB4, CNTNAP2, DLG1, NRCAM, NRXN2, RAPGEF6, RPL14, , RPL15, RPS23, SGCE |
| CD22 | Multi-organism metabolic process | A metabolic process - chemical reactions and pathways, including anabolism and catabolism, by which living organisms transform chemical substances - which involves more than one organism. | 5 / 138 | 3.10E-06 | 2.00E-03 | EIF3L, MAP2K6, RPL14, RPL15, RPS23 |
| CD20 | Muscle contraction | A process in which force is generated within muscle tissue, resulting in a change in muscle geometry. Force generation involves a chemo-mechanical energy conversion step that is carried out by the actin/myosin complex activity, which generates force through ATP hydrolysis. | 4 / 233 | 1.05E-04 | 4.79E-02 | TTN, MYH3, CACNA1H, SCN5A |
| CD20 | Muscle organ development | The process whose specific outcome is the progression of the muscle over time, from its formation to the mature structure. The muscle is an organ consisting of a tissue made up of various elongated cells that are specialized to contract and thus to produce movement and mechanical work. | 5 / 277 | 1.06E-05 | 2.57E-02 | TTN, CACNA1H, MYH3, SGCE, FOXH1 |
| CD20 | Muscle structure development | The progression of a muscle structure over time, from its formation to its mature state. Muscle structures are contractile cells, tissues or organs that are found in multicellular organisms. | 5 / 432 | 8.81E-05 | 4.79E-02 | FOXH1, TTN, MYH3, CACNA1H, SGCE |
| CD20 | Muscle system process | A organ system process carried out at the level of a muscle. Muscle tissue is composed of contractile cells or fibers. | 5 / 282 | 1.16E-05 | 2.57E-02 | TTN, CACNA1H, MYH3, SGCE, SCN5A |
| CD30 | Negative regulation of nitrogen compound metabolic process | Any process that stops, prevents, or reduces the frequency, rate or extent of the chemical reactions and pathways involving nitrogen or nitrogenous compounds. | 15 / 1517 | 1.87E-05 | 1.38E-02 | ADAR, ATN1, BAHD1, BCL7A, CELA1, NANOS1, NCOA2, NKX3-1, NOC2L, PDGFB, PTCH1, TMBIM6, TP63, TRIM22, ZNF93 |
| CD30 | Negative regulation of protein complex assembly | Any process that stops, prevents, or reduces the frequency, rate or extent of protein complex assembly. | 5 / 107 | 1.22E-05 | 1.38E-02 | CLIP3, KANK1, PFN2, VILL, XAF1 |
| CD30 | Negative regulation of protein polymerization | Any process that stops, prevents, or reduces the frequency, rate or extent of the process of creating protein polymers. | 4 / 55 | 1.65E-05 | 1.38E-02 | CLIP3, KANK1, PFN2, VILL |
| CD30 | Negative regulation of response to stimulus | Any process that stops, prevents, or reduces the frequency, rate or extent of a response to a stimulus. Response to stimulus is a change in state or activity of a cell or an organism (in terms of movement, secretion, enzyme production, gene expression, etc.) as a result of a stimulus. | 15 / 1360 | 5.17E-06 | 1.38E-02 | ADAR, DHX58, IRS1, KANK1, NKX3-1, NOC2L, PDGFB, PSMB4, PTCH1, PXDN, SH3BP4, TMBIM6, TNFAIP6, TP63, TXNDC12 |
| CD20 | Organ morphogenesis | Morphogenesis of an organ. An organ is defined as a tissue or set of tissues that work together to perform a specific function or functions. Morphogenesis is the process in which anatomical structures are generated and organized. Organs are commonly observed as visibly distinct structures, but may also exist as loosely associated clusters of cells that work together to perform a specific function or functions. | 7 / 841 | 2.57E-05 | 3.80E-02 | TTN, SCN5A, FOXH1, TP63, DLG5, MEGF8, CELA1 |
| CD22 | Organonitrogen compound metabolic process | The chemical reactions and pathways involving organonitrogen compound. | 13 / 1796 | 5.38E-06 | 2.98E-03 | DLG1, EIF3L, MSH2, NDUFB1, PDE8A, PGLYRP2, PHOSPHO1, RFK, RNF180, RPL14, RPL15, RPS23, TYMP |
| CD22 | Organophosphate metabolic process | The chemical reactions and pathways involving organophosphates, any phosphate-containing organic compound. | 9 / 885 | 1.26E-05 | 5.62E-03 | DLG1, MSH2, NDUFB1, NUDT5, PDE8A, PDGFB, PHOSPHO1, RFK, TYMP |
| CD30 | Organophosphate metabolic process | The chemical reactions and pathways involving organophosphates, any phosphate-containing organic compound. | 11 / 885 | 3.46E-05 | 1.54E-02 | ABHD14B, IRS1, ISYNA1, ME1, NDUFB1, PANK3, PDE8A, PDGFB, PHOSPHO1, PI4K2A, , SMPD3 |
| CD22 | Peptidyl tyrosine modification | The modification of peptidyl-tyrosine. | 5 / 186 | 1.32E-05 | 5.62E-03 | EPHB4, MAP2K6, PDGFB, TNK2, TTN |
| CD22 | Phosphate-containing compound metabolic process | The chemical reactions and pathways involving the phosphate group, the anion or salt of any phosphoric acid. | 15 / 1977 | 5.03E-07 | 1.94E-03 | DLG1, DUSP15, EPHB4, IKBKAP, MAP2K6, MSH2, NDUFB1, NUDT5, PDE8A, PDGFB, PHOSPHO1, RFK, TNK2, TTN, TYMP |
| CD30 | Phosphate-containing compound metabolic process | The chemical reactions and pathways involving the phosphate group, the anion or salt of any phosphoric acid. | 17 / 1977 | 2.99E-05 | 1.53E-02 | ABHD14B, IRS1, ISYNA1, MAP2K6, ME1, NDUFB1, OBSCN, PANK3, PDE8A, PDGFB, PHOSPHO1, PI4K2A, PRKD3, PSMB4, RIPK3, SMPD3, TNK2 |
| CD20 | Polarized epithelial cell differentiation | The process in which a relatively unspecialized cell acquires specialized features of a polarized epithelial cell. The polarized epithelial cell can be any of the cells within an epithelium where the epithelial sheet is oriented with respect to the planar axis. | 2 / 12 | 7.00E-05 | 4.79E-02 | TP63, DLG5 |
| CD30 | Positive regulation of transport | Any process that activates or increases the frequency, rate or extent of the directed movement of substances (such as macromolecules, small molecules, ions) into, out of or within a cell, or between cells, by means of some agent such as a transporter or pore. | 12 / 936 | 1.11E-05 | 1.38E-02 | CASP1, CLIP3, IRS1, KANK1, MAP2K6, NKX3-1, PDGFB, PTCH1, RAB3A, SMPD3, TP63, TRPM4 |

| | | | | | |
|---|---|---|---|---|---|
| CD30 | Regulation of intracellular signal transduction | Any process that modulates the frequency, rate or extent of intracellular signal transduction. | 16 / 1656 | 1.28E-05 | 1.38E-02 | ANKRD54, CASP1, DHX58, IRS1, KANK1, MAP2K6, NKX3-1, NOC2L, OBSCN, PDGFB, RIPK3, SH3BP4, TMBIM6, TP63, TRIM22, TXNDC12 |
| CD22 | Regulation of protein modification process | Any process that modulates the frequency, rate or extent of the covalent alteration of one or more amino acid residues within a protein. | 13 / 1710 | 3.16E-06 | 2.00E-03 | DLG1, DUSP15, GADD45B, GPR183, IKBKAP, MAP2K6, NKX3-1, PDGFB, PKIG, RNF180, TADA2A, TNK2, TTN |
| CD30 | Ribonucleoprotein complex biogenesis | A cellular process that results in the biosynthesis of constituent macromolecules, assembly, and arrangement of constituent parts of a complex containing RNA and proteins. Includes the biosynthesis of the constituent RNA and protein molecules, and those macromolecular modifications that are involved in synthesis or assembly of the ribonucleoprotein complex. | 8 / 440 | 3.11E-05 | 1.53E-02 | ADAR, EIF3L, NOC2L, RPL3, RPS17, RPS23, SART1, SENP3 |

1 – Biological processes pathways as defined by Gene Ontology
2 – The overlap ratio $k/K$ is show the number of genes in the overlap ($k$), by the total number of genes in the gene set ($K$)
3 – Unadjusted p-value calculated from the hypergeometric probability distribution
4 – Adjusted p-value using the Benjamini-Hochberg procedure with FDR

**Supplement 5. Table S2. Molecular pathways involved in CD-related gene sets.**

| CD Predictor | Gene Set Name [1] | Gene Set Description | $k/K$ [2] | p-value [3] | FDR q-value [4] | Genes contained within set overlap |
|---|---|---|---|---|---|---|
| CD30 | Adenosine deaminase activity | Catalysis of the reaction: adenosine + H2O = inosine + NH3. | 2 / 11 | 4.04E-04 | 3.39E-02 | ADAR, ZBP1 |
| CD22 | Adenyl nucleotide binding | Interacting selectively and non-covalently with adenyl nucleotides, any compound consisting of adenosine esterified with (ortho)phosphate. | 12 / 1514 | 5.20E-06 | 2.34E-03 | ACSL6, ABCB4, BAG2, CHD9, DDX54, EPHB4, MAP2K6, MSH2, RFK, TTN, TRPM4, TNK2 |
| CD30 | Adenyl nucleotide binding | Interacting selectively and non-covalently with adenyl nucleotides, any compound consisting of adenosine esterified with (ortho)phosphate. | 17 / 1514 | 9.05E-07 | 2.04E-04 | ACSL6, BBS12, DHX58, DYNC1LI1, KATNAL1, MAP2K6, ME1, OBSCN, PANK3, PFN2, PI4K2A, PMS1, PRKD3, QARS, RIPK3, TNK2, TRPM4 |
| CD20 | Calmodulin binding | Interacting selectively and non-covalently with calmodulin, a calcium-binding protein with many roles, both in the calcium-bound and calcium-free states. | 5 / 179 | 1.27E-06 | 1.14E-03 | SCN5A, USP6, TTN, MYH3, MARCKSL1 |
| CD22 | Enzyme binding | Interacting selectively and non-covalently with any enzyme. | 14 / 1737 | 6.27E-07 | 5.65E-04 | ACSL6, CEP68, CNTNAP2, DLG1, FBL, IKBKAP, MAP2K6, MSH2, NKX3-1, RAPGEF6, RNF180, SGSM1, TRAF3, TTN |
| CD22 | Kinase activity | Catalysis of the transfer of a phosphate group, usually from ATP, to a substrate molecule. | 7 / 842 | 4.20E-04 | 3.64E-02 | ACSL6, ABCB4, ATN1, BAG2, CEP68, EIF3L, IKBKAP |
| CD30 | Macromolecular complex binding | Interacting selectively and non-covalently with any macromolecular complex. | 17 / 1399 | 3.05E-07 | 1.43E-04 | ANKRD54, BAHD1, CLIP3, GNA11, GNAL, IRS1, KATNAL1, KLHL17, MAPRE3, NCOA2, NOC2L, PDGFB, PI4K2A, PTCH1, RIPK3, TADA2A, TP63 |
| CD22 | Molecular function regulator | A molecular function that modulates the activity of a gene product or complex. Examples include enzyme regulators and channel regulators. | 11 / 1353 | 1.06E-05 | 3.18E-03 | BAG2, DLG1, IKBKAP, NRXN2, NKX3-1, PXDN, PDGFB, PKIG, RAPGEF6, SGSM1, TNK2 |
| CD30 | Molecular function regulator | A molecular function that modulates the activity of a gene product or complex. Examples include enzyme regulators and channel regulators. | 14 / 1353 | 2.21E-05 | 3.31E-03 | ANKRD54, CASP1, IRS1, NCOA2, NKX3-1, NRXN2, OBSCN, PDGFB, PFN2, PXDN, RAB3A, SH3BP4, TMBIM6, TNK2 |
| CD30 | Nucleic acid binding transcription factor activity | Interacting selectively and non-covalently with a DNA or RNA sequence in order to modulate transcription. The transcription factor may or may not also interact selectively with a protein or macromolecular complex. | 12 / 1199 | 1.20E-04 | 1.47E-02 | HOXC4, NKX3-1, STAT2, TADA2A, TP63, TRIM22, ZNF134, ZNF420, ZNF561, ZNF606, ZNF79, ZNF93 |
| CD22 | Platelet-derived growth factor receptor binding | Interacting selectively and non-covalently with the platelet-derived growth factor receptor. | 2 / 15 | 2.65E-04 | 2.65E-02 | PDGFB, TYMP |
| CD30 | Protein complex binding | Interacting selectively and non-covalently with any protein complex (a complex of two or more proteins that may include other nonprotein molecules). | 14 / 935 | 3.17E-07 | 1.43E-04 | ANKRD54, CLIP3, GNA11, GNAL, IRS1, KATNAL1, KLHL17, MAPRE3, NCOA2, NOC2L, PDGFB, PI4K2A, PTCH1, RIPK3 |
| CD22 | Protein domain-specific binding | Interacting selectively and non-covalently with a specific domain of a protein. | 7 / 624 | 6.74E-05 | 1.52E-02 | ATN1, CEP68, DLG1, KLHL17, NKX3-1, RNF2, TNK2 |
| CD30 | Protein domain-specific binding | Interacting selectively and non-covalently with a specific domain of a protein. | 10 / 624 | 9.32E-06 | 1.68E-03 | ATN1, HOXC4, IRS1, KHDRBS3, KLHL17, LIN7B, NCOA2, NKX3-1, TNK2, TP63 |
| CD22 | Protein tyrosine kinase activity | Catalysis of the reaction: ATP + a protein tyrosine = ADP + protein tyrosine phosphate. | 4 / 176 | 1.94E-04 | 2.19E-02 | EPHB4, MAP2K6, TTN, TNK2 |
| CD22 | Ribonucleotide binding | Interacting selectively and non-covalently with a ribonucleotide, any compound consisting of a ribonucleoside that is esterified with (ortho)phosphate or an oligophosphate at any hydroxyl group on the ribose moiety | 11 / 1860 | 1.85E-04 | 2.19E-02 | ACSL6, ABCB4, CHD9, DDX54, EPHB4, MAP2K6, MSH2, RFK, TTN, TRPM4, TNK2 |

| | | | | | |
|---|---|---|---|---|---|
| CD30 | Ribonucleotide binding | Interacting selectively and non-covalently with a ribonucleotide, any compound consisting of a ribonucleoside that is esterified with (ortho)phosphate or an oligophosphate at any hydroxyl group on the ribose moiety. | 19 / 1860 | 8.02E-07 | 2.04E-04 | ACSL6, BBS12, DHX58, DYNC1LI1, GNA11, GNAL, KATNAL1, MAP2K6, ME1, OBSCN, PANK3, PI4K2A, PMS1, PRKD3, QARS, RAB3A, RIPK3, TNK2, TRPM4 |
| CD30 | RNA binding | Interacting selectively and non-covalently with an RNA molecule or a portion thereof. | 14 / 1598 | 1.30E-04 | 1.47E-02 | ADAR, DHX58, DYNC1LI1, EIF3L, KHDRBS3, NANOS1, NOC2L, QARS, RPL3, RPS17, RPS23, SART1, ZBP1, ZMAT3 |
| CD22 | Signal transducer activity | Conveys a signal across a cell to trigger a change in cell function or state. A signal is a physical entity or change in state that is used to transfer information in order to trigger a response. | 11 / 1731 | 9.90E-05 | 1.78E-02 | EPHB4, GPR183, IKBKAP, LGR6, HLA-DQB1, MAP2K6, NRXN2, NKX3-1, PGLYRP2, PLXNA3, TRAF3 |
| CD22 | Structural molecule activity | The action of a molecule that contributes to the structural integrity of a complex or assembly within or outside a cell. | 7 / 732 | 1.80E-04 | 2.19E-02 | DLG1, KLHL17, PXDN, RPL14, RPL15, RPS23, TTN |
| CD30 | Transcription factor activity protein binding | Interacting selectively and non-covalently with any protein or protein complex (a complex of two or more proteins that may include other nonprotein molecules), in order to modulate transcription. A protein binding transcription factor may or may not also interact with the template nucleic acid (either DNA or RNA) as well. | 8 / 588 | 2.30E-04 | 2.30E-02 | ATN1, NCOA2, NOC2L, PSMC3IP, RIPK3, TADA2A, TP63, TRIM22 |

1 – Molecular functions pathways as defined by Gene Ontology
2 – The overlap ratio *k*/*K* is show the number of genes in the overlap (*k*), by the total number of genes in the gene set (*K*)
3 – Unadjusted p-value calculated from the hypergeometric probability distribution
4 – Adjusted p-value using the Benjamini-Hochberg procedure with FDR